\documentclass{aa}  

\usepackage{graphicx}
\usepackage{txfonts}
\usepackage{threeparttable} 

\usepackage{natbib,twoopt}
\bibpunct{(}{)}{;}{a}{}{,}             

\begin{document}

   \title{Conditions for radiative zones in the molecular hydrogen envelope of Jupiter and Saturn: The role of alkali metals}

   \author{L. Siebenaler
          \inst{1}
          ,
          Y. Miguel\inst{1}\fnmsep\inst{2},
          S. de Regt \inst{1}
          \and
          T. Guillot\inst{3}
          }

   \institute{Leiden Observatory, University of Leiden,
              Einsteinweg 55, 2333CA Leiden, The Netherlands \\
              \email{siebenalerl@strw.leidenuniv.nl}
         \and
             SRON Netherlands Institute for Space Research , Niels Bohrweg 4, 2333CA Leiden, the Netherlands 
        \and
             Observatoire de la Côte d’Azur, Boulevard de l’Observatoire CS 34229, F 06304 NICE Cedex 4, France
             }

   \date{Received 3 November, 2024; accepted 18 December, 2024}

 
  \abstract
   {Interior models of gas giants in the Solar System traditionally assume a fully convective molecular hydrogen envelope. However, recent observations from the Juno mission suggest a possible depletion of alkali metals in Jupiter's molecular hydrogen envelope, indicating that a stable radiative layer could exist at the kilobar level. Recent studies propose that deep stable layers help reconcile various Jupiter observations, including its atmospheric water and CO abundances and the depth of its zonal winds. However, opacity tables used to infer stable layers are often outdated and incomplete, leaving the precise molecular hydrogen envelope composition required for a deep radiative zone uncertain.}
   {In this paper, we determine atmospheric compositions that can lead to the formation of a radiative zone at the kilobar level in Jupiter and Saturn today.
   }
   {We computed radiative opacity tables covering pressures up to $10^5$ bar, including the most abundant molecules present in the gas giants of the Solar System, as well as contributions from free electrons, metal hydrides, oxides, and atomic species, using the most up-to-date line lists published in the literature. These tables were used to calculate Rosseland-mean opacities for the molecular hydrogen envelopes of Jupiter and Saturn, which were then compared to the critical mean opacity required to maintain convection.
   
   }
   {We find that the presence of a radiative zone is controlled by the existence of K, Na, and NaH in the atmosphere of Jupiter and Saturn. For Jupiter, the elemental abundance of K and Na must be less than $\sim 10^{-3}$ times solar to form a radiative zone. In contrast, for Saturn, the required abundance for K and Na is below $\sim 10^{-4}$ times solar.
   }
   {}

\titlerunning{Conditions for radiative zones in Jupiter and Saturn}
\authorrunning{Siebenaler et al.}
\maketitle
%

\section{Introduction}

\begin{figure}
   \centering
   \includegraphics[width=0.4\textwidth]{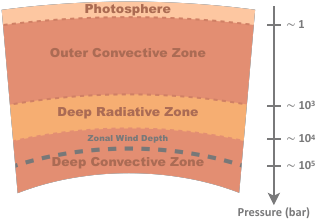}
   \caption{Schematic view of Jupiter's molecular hydrogen envelope. Low pressures ($\lesssim 1$ bar) correspond to the radiative photosphere, which based on chemistry codes is characterised by nitrogen-bearing clouds. Below is an outer convective zone that sits on top of a potential radiative layer. The radiative layer extends from $\sim 10^3 - 10^4$ bar, at which refractory species condense. At pressures $\gtrsim 10^4$ bar, the atmosphere is convective again. We only show several condensing species for illustrative purposes. The structure of Saturn would be similar, although with a different zonal wind depth.}
    \label{fig:Jupiter_envelope}
    \end{figure}

Traditionally, interior models of gas giants assume convection as the dominant heat transport mechanism in the hydrogen-dominated envelope, which is driven by their high internal energy. As a consequence, the planet's thermal profile is typically described by an adiabat, with the envelope expected to have a uniform composition due to convective mixing. However, \cite{Guillot_1994} demonstrated that the assumption of a fully convective interior may not always hold. At temperatures of $\sim 2000$ K, it was found that the giant planets in the Solar System exhibit a decrease in opacity, which allows radiation to become the dominant heat transport mechanism and means that a stable radiative zone might develop. The presence of such a radiative zone can have significant effects on the planet. For example, reduced mixing efficiency across the radiative layer can prevent a uniform composition in the molecular hydrogen envelope, leading to differences between atmospheric and interior models. A sketch of Jupiter's molecular envelope in the presence of a radiative zone is given in Fig. \ref{fig:Jupiter_envelope}. This shows a radiative layer located between $\sim 10^3 - 10^4$ bar, and surrounded by two convective regions. However, this early study by \cite{Guillot_1994} neglected the opacity of alkali metals, which makes a significant contribution to the absorption of photons in this region. More recent opacity calculations \citep{Guillot_2004, Freedman_2008} have shown that alkali opacities can inhibit radiative energy transport and restore convection. Hence, the idea of a stable radiative zone in gas giants was largely dismissed due to the supersolar metallicity measurements of Jupiter's \citep{Mahaffy_2000, Wong_2004} and Saturn's \citep{Fletcher_2012} atmospheres. However, recent findings from the Juno mission \citep{Bolton_2017} have significantly improved our understanding of Jupiter's atmosphere and interior, pointing to the potential presence of radiative zones throughout the molecular hydrogen envelope. Recent interpretation of Juno microwave radiometer (MWR) data suggest that Jupiter's atmosphere exhibits a depletion in alkali metals \citep{Bhattacharya_2023, Aglyamov_2024}, decreasing the opacity around $\sim 2000$ K significantly, which disfavours convection and could restore the original hypothesis by \cite{Guillot_1994}. Additional support for this radiative zone comes from the low atmospheric CO abundance  \citep{Bezard_2002, Bjoraker_2018} and recent MWR measurements of high deep-water abundance \citep{Li_2020}. These observations align with suppressed vertical mixing, which is consistent with the presence of a deep radiative zone \citep{Cavalie_2023}. Moreover, an inverted $Z$ gradient produced by a radiative zone could help resolve the tension between atmospheric constraints and interior models \citep{Howard_2023}. This tension involves the subsolar to solar heavy element abundances in Jupiter’s molecular hydrogen envelope predicted by interior models, in contrast with the supersolar abundances observed by the Juno mission and Galileo probe. \cite{Muller_2024} demonstrate that such an inverted $Z$ gradient can be maintained by a radiative zone, which prevents downward mixing between the atmosphere and envelope, although this requires an extremely small diffusion coefficient close to molecular diffusivity. Further support for stable layers within the molecular hydrogen envelope comes from the inferred depth of Jupiter's zonal winds  \citep{Christensen_2024} and the characteristics of Jupiter's dipolar magnetic field \citep{Moore_2022}. However, the stable layers needed to explain these observations are located deeper in the envelope, at depths of $\gtrsim 2000$ km, compared to the shallower radiative layers produced by alkali metal depletion.

Despite recent results from Juno, the existence of a radiative zone on Jupiter is still unclear. This is related to the fact that most studies on the radiative zone rely on outdated and incomplete opacity data, and it is unclear how much alkalis are needed to prevent a radiative zone. In this work, we construct new radiative opacity tables to assess the existence of a radiative layer on Jupiter and Saturn, the goal being to determine the atmospheric compositions needed to prevent a radiative layer. In Sect. \ref{sec:Method}, we explain how we calculate the atmospheric composition of the planets, and the method of determining a stable radiative layer. We also present the sources of opacity used in this work and how they were computed. In Sect. \ref{sec:Results}, we calculate mean opacities for Jupiter and Saturn and determine the conditions needed for a radiative zone. Moreover, we construct non-adiabatic thermal profiles that include the presence of a radiative zone. Section \ref{sec:Discussion} is devoted to discussion points, and finally, in Sect. \ref{sec:Conclusions}, we give our conclusions.

\section{Method} \label{sec:Method}

\subsection{Atmospheric composition} \label{sec:Atmospheric_Composition}

To assess the existence of a stable radiative layer in the molecular hydrogen envelope of gas giants, we need to assume an atmospheric composition. In this section, we outline the general knowledge of the atmospheric composition of Jupiter and Saturn, and explain how the levels inaccessible to observations are modelled.

Atmospheric remote sensing from missions like Juno, Galileo, and Cassini has provided detailed insights into the atmospheric compositions of Jupiter and Saturn. These measurements generally show that both planets' atmospheres are enriched in heavy elements\footnote{In this work, we refer to heavy elements and metals as elements heavier than H and He.}, with Jupiter exhibiting about three times solar enrichment and Saturn around eight times solar. For Jupiter, we have additional in situ measurements by the Galileo probe down to the 20 bar level, which revealed a slight depletion in helium, $Y_{\rm atm} = 0.238 \pm 0.005$ ($\sim 0.8$ times solar; \citealt{von_Zahn_1998, Niemann_1998}). This suggests that some helium rain has occurred on Jupiter. In contrast, Saturn's atmospheric helium abundance remains uncertain due to the lack of in situ measurements. Remote measurements by Voyager ($Y_{\rm atm} = 0.18 - 0.25$; \citealt{Conrath_2000}), Cassini ($Y_{\rm atm} = 0.158 - 0.217$; \citealt{Koskinen_2018}), and Cassini Grand Finale Tour  ($Y_{\rm atm} = 0.075 - 0.130$; \citealt{Waite_2018}) suggest that Saturn's atmosphere shows a greater helium depletion compared to Jupiter. This is consistent with predictions from evolution models \citep{Nettelmann_2015, Mankovich_2020, Howard_2024}, which suggest a more pronounced helium rain process in Saturn due to its colder interior. Helium constitutes the second-largest mass fraction in gas giants and its abundance influences the temperature gradients in the deep atmosphere, which in turn controls the chemistry of the planet.
Since pressures above $\sim 300$ bar are inaccessible to current space missions, modelling the atmospheric composition at these depths requires one to assume a thermal profile. In this work, we model the temperature-pressure profile (TP profile) of the planet as a dry adiabat, with a 1-bar temperature fixed to the value given in Table \ref{table:fundamental_properties}. The adiabatic gradient was computed using the \cite{Chabrier_2019} equation of state for dense H-He mixtures and the non-ideal mixing terms from \cite{Howard_2023} as

\begin{table}[t]
\small
\centering
 \caption{Fundamental properties of Jupiter and Saturn.}
 \begin{tabular}{c @{\hskip 0.3cm }  c  @{\hskip 0.3cm }  c} 
 \hline
 \hline
& Jupiter  & Saturn \\ [0.01ex] 
  \noalign{\smallskip}
    \hline
    \noalign{\smallskip}
    Mass$^{\rm a}$ ($10^{24}$ kg) & $1898.16\pm 0.15$ & $568.463\pm 0.016$ \\
    Equatorial radius$^{\rm b}$ (km) & $71492\pm 4$ & $60268\pm 4$ \\
    Absorbed power$^{\rm c}$ (W m$^{-2}$) & $6.613 \pm 0.160$ & $2.04\pm 0.17$ \\
    Internal energy flux$^{\rm c}$ (W m$^{-2}$) & $7.485\pm 0.163$ & $2.84\pm 0.20$ \\
    Total energy flux (W m$^{-2}$) & $14.098\pm 0.031$ & $4.88\pm 0.11$ \\
    1-bar temperature$^{\rm d}$ (K) & $170 \pm 5$ & $135 \pm 5$ \\       
  \hline
\end{tabular}
\begin{flushleft}
\tiny{ $^{\rm a}$\cite{Campbell_1985} for Jupiter; \cite{Jacobson_2006} for Saturn  } \\
\tiny{ $^{\rm b}$\cite{Lindal_1981} for Jupiter; \cite{Lindal_1985} for Saturn  } \\
\tiny{ $^{\rm c}$\cite{Li_2018} for Jupiter; \cite{Wang_2024} for Saturn  } \\
\tiny{ $^{\rm d}$\cite{Seiff_1998} and \cite{Gupta_2022} for Jupiter; \cite{Lindal_1985} for Saturn  }
\end{flushleft}

 \label{table:fundamental_properties}

\end{table}

\begin{equation}
    \nabla_{\rm ad} = \frac{(1 - Y_{\rm atm}) S_{\rm H} \bigg( \frac{\partial \textrm{log} S_{\rm H}}{\partial \textrm{log} P}\bigg)_T + Y_{\rm atm} S_{\rm He} \bigg( \frac{\partial \textrm{log} S_{\rm He}}{\partial \textrm{log} P}\bigg)_T + \Delta S \bigg( \frac{\partial \textrm{log} \Delta S}{\partial \textrm{log} P}\bigg)_T}{(1 - Y_{\rm atm}) S_{\rm H} \bigg( \frac{\partial \textrm{log} S_{\rm H}}{\partial \textrm{log} T}\bigg)_P + Y_{\rm atm} S_{\rm He} \bigg( \frac{\partial \textrm{log} S_{\rm He}}{\partial \textrm{log} T}\bigg)_P + \Delta S \bigg( \frac{\partial \textrm{log} \Delta S}{\partial \textrm{log} T}\bigg)_P} .
\end{equation}

\begin{figure}[t]
   \centering
   \includegraphics[width=0.45\textwidth]{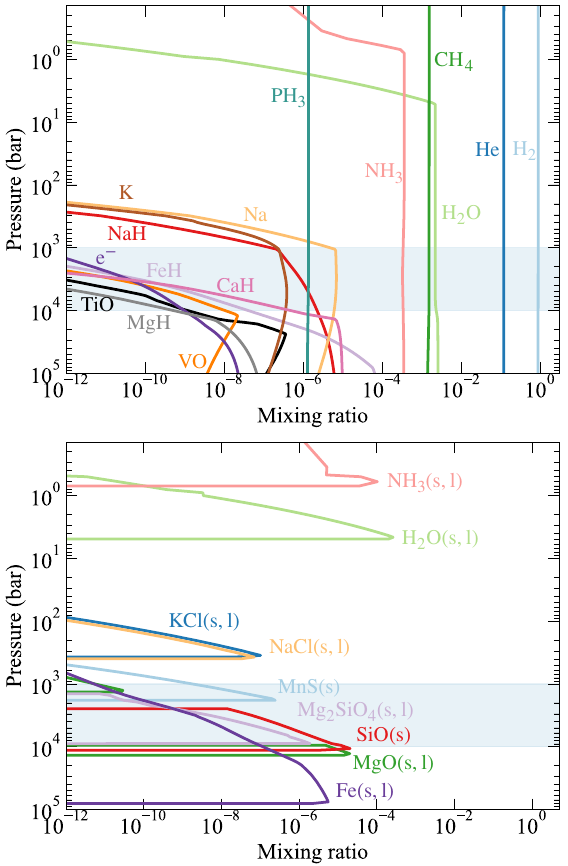}
   \caption{Chemistry of Jupiter assuming an abundance of three times solar for all heavy elements. The upper panel applies to the gas chemistry, while the lower panel gives the condensate chemistry. The approximate pressure range for the potential radiative zone is shown by the shaded blue area.}
    \label{fig:Jupiter_chemistry}
    \end{figure}

\noindent Here, $S_{\rm H}$ and $S_{\rm He}$ denote the specific entropy of pure hydrogen and helium, respectively, and $\Delta S = Y_{\rm atm} (1- Y_{\rm atm}) S_{\rm mix}$ is the entropy mixing term. We adopted a helium abundance of $Y_{\rm atm} = 0.238$ for Jupiter, while for Saturn we used $Y_{\rm atm} = 0.204$. For pressures below 1 bar, we used temperatures derived from in situ measurements by the Galileo probe for Jupiter \citep{Seiff_1998} and from Voyager radio-occultation data for Saturn \citep{Lindal_1985}.  We computed the chemistry of the planets along the resulting TP profiles using the equilibrium chemistry code \texttt{FastChem} \citep{Kitzmann_2023}. Due to the high temperatures ($\sim 1000 - 2000$ K) and pressures ($\sim 10^3 - 10^4$ bar) relevant to this study, it is reasonable to assume that chemical timescales are shorter than dynamical ones, justifying the use of equilibrium chemistry. To model the removal of species in the gas phase into the condensate phase and their subsequent settling into cloud layers, we adopted the rainout approximation \citep{Lodders_2002}. We show the gas and condensate chemistry of Jupiter using this approach in Fig. \ref{fig:Jupiter_chemistry}. In these calculations, the abundance of all heavy elements was set to three times the solar value of \cite{Asplund_2021}. As was expected, the atmosphere is dominated by H$_2$ and He gas, with other molecules such as H$_2$O, CH$_4$, and NH$_3$  also abundant throughout in gas form. At pressures $\gtrsim 1000$ bar, alkali species such as K, Na, and NaH become abundant, which aligns with the onset of the stable radiative layer in Jupiter determined by \cite{Guillot_1994}. 
Deeper in the atmosphere, metal hydrides and oxides form, and the abundance of free electrons increases. Condensation also plays a key role. Based on chemistry models, a water crystalline layer is expected around the 1-bar level of Jupiter. Around 100 bar, alkali species condense out of the atmosphere, which causes them to only become abundant in gas form at deeper levels. At even higher pressures, in the region of the potential radiative zone, refractory species such as silicates and iron condense, forming cloud layers, as is shown in the lower panel of Fig. \ref{fig:Jupiter_chemistry}. Saturn’s chemistry is quite similar to Jupiter’s, but its slightly cooler temperature causes elements like K and Na, and metal hydrides and oxides, to form at higher pressures. Similarly, condensation occurs at higher pressures, and the mixing ratios of most species are greater due to Saturn’s higher metallicity.

\subsection{Heat transport of giant planets} \label{sec:heat_transport}

In the H$_2$-envelope, the main mechanisms to transport internal energy are radiation and convection. In this work, we use the Schwarzschild stability criterion \citep{Schwarzschild_1958} to determine the dominant mechanism. Deep in the atmosphere, where the optical depth is high, the diffusion approximation must hold. Under these conditions, the Rosseland-mean opacity, $\kappa_{\rm R}$, controls the efficiency  of radiative heat transport (e.g. \citealt{Rybicki_1986}). It is defined as 

\begin{equation}\label{rosseland_opacity}
    \frac{1}{\kappa_{\rm R}} = \frac{\int_0^\infty \kappa_{\nu}^{-1} \frac{dB_\nu}{dT} d\nu}{\int_0^\infty \frac{dB_\nu}{dT} d\nu},
\end{equation}

\noindent where $\kappa_{\nu}$ is the monochromatic opacity and $B_\nu$ is the Planck function. From the definition of the radiative temperature gradient (e.g. \citealt{Guillot_1994}) and the Schwarzschild stability criterion, it can be shown that if $\kappa_{\rm R}$ of a layer exceeds a critical value,

\begin{equation}\label{eq:critical_opacity}
    \kappa_{\rm crit} = \frac{16 \rm g}{3 P} \Bigg(\frac{T}{T_{\rm eff}} \Bigg)^4 \nabla_{\rm ad}
,\end{equation}

\noindent it becomes convective. Here, $P$ and $T$ are the pressure and temperature of the layer, and $\rm g$ is the gravitational acceleration, which is assumed to be independent of depth in this study. $T_{\rm eff}$ corresponds to the effective temperature of a layer and is derived from the total luminosity transported at that level. At high optical depths, all the irradiated flux has been absorbed so that $T_{\rm eff}$ is given by the internal energy flux alone. Since we adopted the diffusion approximation to describe the radiative heat transport, we assume that a layer is always optically thick when computing $T_{\rm eff}$. As a result, we used $T_{\rm eff} = 107.19 \pm 0.57$ K  for Jupiter, and  $T_{\rm eff} = 84.13 \pm 1.48$ K for Saturn, which was derived from the internal energy fluxes given in Table \ref{table:fundamental_properties}. We realise that our treatment of $T_{\rm eff}$ may not be valid at pressures $\lesssim 1$ bar, where the optical depth is low and the irradiated flux needs to be accounted for, which causes $T_{\rm eff}$ to increase slightly. In addition, at low optical depth the diffusion approximation breaks down, which implies that the application of $\kappa_{\rm R}$ is invalid. In this regime, the radiative transfer equation has to be solved to determine the radiative temperature gradient. However, the main focus of this study is the kilobar regime, where our treatment of $T_{\rm eff}$ and Eq. (\ref{eq:critical_opacity}) is valid. From Eq. (\ref{eq:critical_opacity}), it becomes clear that the presence of a radiative zone in gas giant atmospheres is controlled by opacities. We shall describe our treatment of radiative opacity calculations in the following section.

\subsection{Opacity calculations}

In this section, we describe the calculations of opacities used in this work. In the region of a potential radiative zone ($\sim 1000 - 2000$ K), the Planck function peaks around  1 $\rm \mu m$. Consequently, we considered a spectral domain from 0.1 - 200 $\rm \mu  m$ to compute opacities. Within this range, absorption due to molecules, atoms, collision-induced absorption (CIA), free electrons, and scattering all contribute to $\kappa_{\rm R}$. In general, we calculated the absorption cross-sections over a grid of temperatures and pressures, which covers a temperature range from 100 to 5000 K with 50 linearly spaced points, and a pressure range from $10^{-1}$ to $10^5$ bar with 26 points evenly spaced in log. The radiative opacities used in this study, along with tables of $\kappa_{\rm R}$, are publicly available in the Zenodo repository\footnote{\url{https://doi.org/10.5281/zenodo.14507606}}.

\subsubsection{Atomic line absorption} \label{sec:atomic_lines}
This study considers the opacities of neutral atoms: Ca, Fe, Mg, Mn, Ni, Ti, V, Na, and K. We focus on atomic species that appear as condensates in the equilibrium chemistry code, ensuring their correct removal from the gas phase. While additional atomic species could be considered, their opacities are far less significant compared to molecular opacities.
We calculated the absorption cross-sections of the non-alkali atoms in a uniform way using data from NIST \citep{NIST_2001} and the Vienna Atomic Line Data (VALD - \citealt{Ryabchikova_2017}). We implemented our own atomic opacity calculator and shall now outline the main equations involved in these calculations.

Atomic (and molecular) absorption lines are described by an integrated line strength, $S$, which can be calculated assuming local thermodynamic equilibrium. For a transition from a level, $i$, to a higher level, $j$, we have (in cm g$^{-1}$)

\begin{equation}
S = \frac{\pi \textrm{e}^2 g_i  f_{ij}}{\rm m_e c^2 \emph{Q(T)}} e^{- \frac{\rm hc}{\rm k_B \emph{T}} E_i } \Bigg(   1 - e^{- \frac{\rm hc}{\rm k_B \emph{T}} (E_j - E_i)}\Bigg),
\end{equation}

\noindent where, $E_i$ and $E_j$ are the excitation energies in cm$^{-1}$ of the $i^{\rm th}$ and $j^{\rm th}$ level. The factors, $g_i$ and $f_{ij}$, correspond, respectively, to the statistical weight of the $i^{\rm th}$ energy level and the oscillator strength of the transition. We  retrieved the product of $g_i f_{ij}$ values from VALD. The partition function of the species is given by $Q(T)$, and we computed it using the statistical weights from NIST.

Absorption lines are broadened due to thermal and pressure effects. Thermal effects produce a Gaussian line profile with a half width at half maximum \citep[HWHM; in cm$^{-1}$;][]{Hill_2013},

\begin{equation}\label{eq:Doppler_width}
    \gamma_{\rm G} = \frac{\nu_0}{\rm c} \sqrt{\frac{2 \textrm{ln(2)k}_{\rm B} T}{m}}, 
\end{equation}

\noindent where $\nu_0$ is the line centre in wavenumber and $m$ is the mass of the species. 

Pressure effects influence the lifetime of an energy state, which in turn specifies the broadening of the absorption line. In general, the most important pressure broadening effect in a dense medium is Van der Waals broadening. Another, less significant effect, is natural broadening as a consequence of the uncertainty principle \citep{Gray_2008}. Both effects introduce a Lorentzian absorption profile, with a combined HWHM (in cm$^{-1}$) given by \citep{Sharp_2007},

\begin{equation} \label{eq:gamma_L_1}
    \gamma_{\rm L} = \frac{1}{4 \pi \rm c} \Bigg[ \gamma_{\rm W} (C_{\rm H_2} N_{\rm H_2} + C_{\rm He} N_{\rm He})
    \Bigg(\frac{T}{10^4 \rm K} \Bigg)^{3/10} + \gamma_N \Bigg],
\end{equation}

\noindent where $\gamma_{\rm W}$ and $\gamma_{\rm N}$ are the Van der Waals and natural broadening coefficients, respectively, and were retrieved from VALD. $N_{\rm H_2}$ and $N_{\rm He}$ are the number densities of H$_2$ and He, respectively. The coefficients, $C_{\rm H_2} = 0.85$ and $C_{\rm He} = 0.42$, account for the different polarisabilities of H$_2$ and He \citep{Kurucz_1979}. It is evident that in Eq. (\ref{eq:gamma_L_1}) we assume that atoms are only affected by perturbation from H$_2$ and He, which are the most abundant perturbers of gas giant atmospheres. For some transitions, Van der Waals broadening coefficients, $\gamma_{\rm W}$, are unavailable. In this case, we define the Lorentzian HWHM (in cm$^{-1}$) as (\citealp{Schweitzer_1996, Sharp_2007}),

\begin{equation} \label{eq:gamma_L_2}
    \gamma_{\rm L} = \frac{1.664461}{2\rm c} \Bigg( \textrm{k}_{\rm B} T  \Bigg)^{3/10} \sum_{p} \Bigg[\Bigg( \frac{1}{m} + \frac{1}{ m_{p}}\Bigg)^{3/10}C_{6, p}^{2/5} N_{p} \Bigg] + \frac{\gamma_{\rm N}}{4\pi \rm c},
\end{equation}

\noindent where $m_p$ and $N_p$ are the mass and number density of the perturber. As in Eq. (\ref{eq:gamma_L_1}), we only considered the most abundant perturbers, H$_2$ and He.  The parameter $C_{6, p}$ is known as the interaction constant for Van der Waals interactions of a specific perturber. In general, for a transition between an energy level, $E_i$, to $E_j$, it is given as (in $\rm cm^6 \ s^{-1}$)

\begin{equation}
    C_{6, p} = %
    1.01 \cdot 10^{-32} \frac{\alpha_{p}}{\alpha_{\rm H}} (Z + 1)^2 \Bigg[ \frac{E_{\rm H}^2}{(E - E_{i})^2} - \frac{E_{\rm H}^2}{(E - E_{j})^2}  \Bigg].
\end{equation}

\noindent Here, $\alpha_{\rm p}$ defines the polarisabilty of the perturber, 
$Z$ is the charge of the absorber (we only consider neutral atoms, and hence we always have $Z = 1$), $E$ is the ionisation potential of the absorber, $\alpha_{\rm H}$ is the polarisability of hydrogen, and $E_{\rm H} = 13.6 \ \rm eV$. For the polarisabilities of H$_2$, and He, we used the values as given in Table 1 in \cite{Schweitzer_1996}. Since absorption lines have contributions from a Gaussian and Lorentzian profile, we have modelled them as Voigt profiles, which we express in terms of the real part of the normalised Faddeeva function \citep{Gandhi_2020}.

Another important parameter when calculating absorption lines is the line wing cut-off, which defines the extent of the line wing on either side from the line centre. While its importance is well known in the community, there exists no first principle theory on how to treat this parameter for atoms and molecules in general. We followed the standard practice of the Exoclime team of truncating the line wings at an arbitrary distance of 100 cm$^{-1}$ from the line centre (e.g. \citealt{Malik_2019, Whittaker_2022}). This is in line with the standard practice procedure for high pressures recommended by \cite{Nezhad_2023}.

A different approach was used for the alkali atoms K and Na, which are very important species for the existence of a radiative layer (see Sect. \ref{sec:opacity_spectra}). Observations of exoplanets and brown dwarfs have revealed that their resonance absorption lines exhibit strong deviations from the usual Voigt profile in their far wings (e.g. \citealt{Burrows_2000}). This is caused by perturbations with other species, in particular H$_2$ and He, where the potential-energy curves describing the interactions differ from a Van der Waals broadening potential. The transitions in question are the ones of the K I resonance lines at $0.770 \ \mu\rm m$ and the Na D resonance lines at $0.589 \ \mu \rm m$. There have been various studies of these line profiles in the past to determine precise interaction potentials (e.g. \citealt{Tsuji_1999, Burrows_2000, Burrows_2003, Allard_2012}). In this work, we use the most recent theoretical calculations of the resonance line profiles perturbed by H$_2$ from \cite{Allard_2016} for K and \cite{Allard_2019} for Na. These calculations are valid up to a H$_2$ number density of $N_{\rm H_2} = 10^{21} \rm \ cm^{-3}$. Hence, there will be a critical partial pressure, above which a different treatment of the resonance line profiles has to be used. For a Jupiter-like atmosphere, in which the H$_2$ volume mixing ratio is $\sim 88 \%$, we can estimate that this critical pressure ranges between $10 - 600$ bar for the temperatures of interest here. Unfortunately, as of this point there exist no theoretical calculations for the resonant line profiles at higher pressures. As a result, we modelled the line profiles again using a Voigt function in a similar way to the other atomic absorption lines; that is, using Eq. (\ref{eq:gamma_L_1}) and the Van der Waals coefficients from VALD. However, one difference from the previous calculations is that we adopted a line wing cut-off of 4500 cm$^{-1}$ for the K and Na resonance lines, as is recommended by \cite{Baudino_2017} to still account for their pseudo-continuum opacity spanning from the optical to the near-infrared. For all the other lines, we still employed a 100 cm$^{-1}$ cut-off. We discuss the consequences of the HWHM calculation and adopted line wing cut-off for the resonance lines in Sect. \ref{sec:cutoff_HFWHM}.

\subsubsection{Molecular and collision-induced absorption}
The absorption spectrum of molecules and CIA is calculated using the opacity calculator \texttt{HELIOS-K} \citep{Grimm_2021}. This allows one to compute absorption cross-sections from line lists and partition functions available from online databases. The opacities are computed at a resolution of $10^{-2}$ cm$^{-1}$. For molecules, we chose the ExoMol database \citep{Tennyson_2016, Tennyson_2024} and used the line lists recommended by their team at the time of this work (May 2024). The only exception is CH$_4$, for which we used the most recent HITEMP line list instead of the ExoMol recommendation, as it covers shorter wavelengths. As for the atomic lines, the molecular lines are approximated by Voigt profiles and we adopted a line wing cut-off of 100 cm$^{-1}$ from the line centre. For the pressure broadening, we used the standard broadening parameters provided by the ExoMol database. The CIA corresponds to a continuum opacity rather than distinct spectral lines, so no assumption about line profiles is necessary. We used the data from HITRAN \citep{Gordon_2022} to calculate them. Tables \ref{table:molecule_opacity} and \ref{table:CIA_opacity} summarise all the molecular and CIA opacities considered in this study. In general, we calculated the absorption cross-sections over the aforementioned temperature-pressure grid. However, it is important to note that the line list of some species (CP, CrH, H$_2$S, NaCl, NH$_3$, TiH, and all CIA) do not extend across this entire temperature range, which explains their specific temperature ranges. To calculate opacities outside of their temperature range, we used the closest available absorption cross-sections in temperature space, without performing any extrapolation. Furthermore, we used the pre-calculated opacity data from the DACE\footnote{\url{https://dace.unige.ch/opacityDatabase/}} database for CH$_4$, CO$_2$, PH$_3$, and SiH$_4$. These are larger molecules with billions of transitions, making the calculation of their absorption cross-sections computationally very expensive. Although DACE uses a different temperature-pressure grid than the one adopted in this work, it covers the relevant conditions in which these molecules significantly contribute to $\kappa_{\rm R}$. For H$_2$O and NH$_3$, which are also larger molecules, we used their available super-line lists to calculate their absorption cross-sections.

\subsubsection{Bound-free and free-free absorption}
At high temperatures ($\gtrsim 2500$ K), bound-free and free-free interactions become relevant sources of opacities. In bound-free processes, a sufficiently energetic photon detaches an electron from an ion, and in free-free processes, a free electron approaches an atom or molecule, altering its potential. Both interactions result in continuous absorption of photons. In our calculations, we included bound-free absorption from H$^{-}$, and free-free absorption from H$_2^{-}$, H$^{-}$, and He$^{-}$. We summarise the reactions, the wavelength coverage of the opacities, and the relevant references in Table \ref{table:BF_FF_opacity}. 

\subsubsection{Rayleigh scattering}
Rayleigh scattering acts as a continuous source of opacity. We included Rayleigh scattering cross-sections of the following species: CO$_2$ (\citealt{Sneep_2005, Thalman_2014}), CO (\citealt{Sneep_2005}), H$_2$ \citep{Cox_2000}, H \citep{Lee_2004}, He (\citealt{Sneep_2005, Thalman_2014}), N$_2$ (\citealp{Sneep_2005, Thalman_2014}), O$_2$ (\citealp{Sneep_2005, Thalman_2014}), and electrons \citep{astropy_2022}.

\subsubsection{Abundance weighted-total opacities}

After computing the cross-sections, we interpolated them for each species to the same wavelength grid of resolution $\Delta \lambda / \lambda = 10000$. We found that this resolution results in converging values of $\kappa_{\rm R}$. To convert the  absorption and scattering cross-sections (in cm$^2$) into opacities (cm$^2$ g$^{-1}$), we multiplied them by the mixing ratio of the species and divided this by the mean molecular weight of the atmospheric layer\footnote{CIA and free-free absorption cross-sections need to be multiplied by the mixing ratios of both species involved in the process.}. 
As is mentioned in Sect. \ref{sec:Atmospheric_Composition}, we computed $\kappa_{\rm R}$ along a dry adiabatic TP profile of Jupiter and Saturn. The temperatures and pressures along these profiles do not exactly match our cross-section grid. Hence, we used bilinear interpolation to match the cross-sections to the temperature-pressure conditions along the profiles. The interpolation was performed in logarithmic space for pressure and linear space in temperature.


\section{Results} \label{sec:Results}
\subsection{Opacity spectra} \label{sec:opacity_spectra}

We now want to identify the dominant sources of opacity at different depths in the atmosphere of gas giants in the Solar System. Assuming a general solar composition for the heavy elements, we focus on various layers along Jupiter's adiabat. The following conclusions can also be applied to Saturn.

\begin{figure}[t!]
   \centering
   \includegraphics[width=0.45\textwidth]{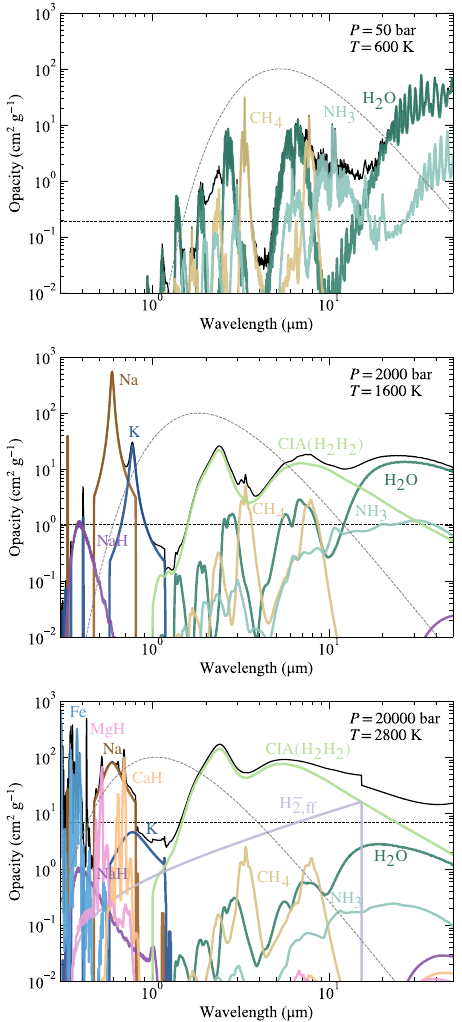}
   \caption{Opacities of Jupiter at different depths, assuming a solar composition in heavy elements. The total opacity of the layer is given in black, and the opacities of individual species are given as coloured curves. The Rosseland-mean opacity, $\kappa_{\rm R}$, of the layer is shown as the dashed black line and the Planck function at the relevant temperature is given as the dashed grey curve.}
    \label{fig:Jupiter_opacity_spectra}
    \end{figure}

Figure \ref{fig:Jupiter_opacity_spectra} shows the total opacity (solid black curves) of our atmospheric Jupiter model at three different depths. The upper panel applies to a pressure of 50 bar, which corresponds to the outer convective region of Jupiter. At these conditions, the most abundant molecules in the gas phase are H$_2$O, CH$_4$, and NH$_3$, all of which have strong absorption features in the infrared. Wavelengths above 1 $\rm \mu m$ contribute the most to $\kappa_{\rm R}$ in this layer, as is highlighted by the Planck curve (dashed grey), which acts as a weighting function to $\kappa_{\rm R}$. Hence, H$_2$O, CH$_4$, and NH$_3$ control the opacity and heat transport in this layer, which is a picture that persists until about $10^3$ bar. At deeper layers, the higher density makes CIA significantly more important. The middle panel in Fig. \ref{fig:Jupiter_opacity_spectra} shows the opacity at a layer of $2 \cdot 10^3$ bar. This reflects the importance of CIA by H$_2$ at these conditions, which now overwhelms the opacity of molecules at longer wavelengths. Below 1 $\rm \mu m$, the absorption of alkali-bearing species, K, Na, and NaH emerges and completely dominates the short-wavelength opacity. We remind the reader that the resonance lines K I and Na D are modelled as Voigt profiles here, due to the absence of a detailed theory for their potential-energy curves describing the interaction with H$_2$ at high pressures. The higher temperature of the layer causes the Planck function to shift to lower wavelengths, which causes the alkali species to make an important contribution to $\kappa_{\rm R}$. In the absence of alkali species, $\kappa_{\rm R}$ of this layer would drop significantly, as no other short-wavelength absorbing species are present in considerable amounts at this depth.
Moreover, because $\kappa_{\rm R}$ is a harmonic mean, the few absorbers in the visible spectrum can impact it much more than the numerous absorbers in the infrared. Other short-wavelength absorbing species only become abundant at layers around $10^4$ bar. This includes metal hydrides (ex. CaH, CrH, FeH, MgH), metal oxides (e.g. CaO, MgO, TiO, VO), bound-free and free-free absorptions due to electrons, and other neutral atomic species (e.g. Ca, Fe, Mg). The lower panel in Fig. \ref{fig:Jupiter_opacity_spectra} gives the opacity spectrum at $2 \cdot 10^4$ bar, and shows the absorption of several species at short wavelengths. Hence, at high pressures a range of species contribute to $\kappa_{\rm R}$, not only alkali gases.

The fact that between $\sim 10^3-10^4$ bar, alkali are the only species capable of absorbing in the visible causes them to control the existence of the radiative layer that was predicted by \cite{Guillot_1994}, who did not include those opacities in their calculations. Its importance was later realised in \cite{Guillot_2004}, in which they found that they could prevent the radiative zone. In the next section, we shall estimate the abundance of K and Na needed to achieve this.

\subsection{Conditions for a radiative zone on Jupiter}

\subsubsection{Deep radiative zone}

\begin{figure}
   \centering
   \includegraphics[width=0.5\textwidth]{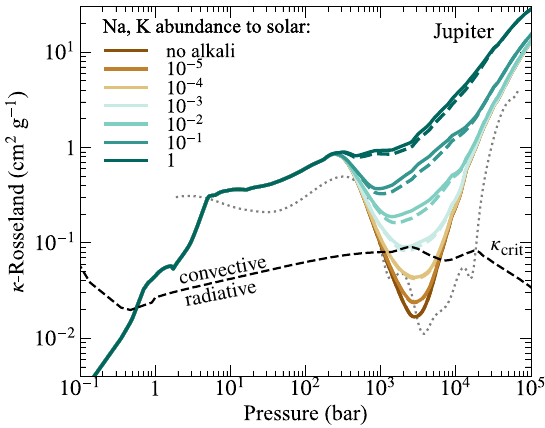}
   \caption{Rosseland-mean opacity, $\kappa_{\rm R}$ (solid and dashed coloured curves), as a function of pressure for atmospheres with varying and Na and K abundances calculated along a Jupiter profile. The abundance of all other heavy elements is three times solar. The solid curves were computed using the Na D and K I HWHM formulation from Eq. (\ref{eq:gamma_L_1}), while the dashed curves use the Na D and K I HWHM formulation from \cite{Allard_2016}, \cite{Allard_2019}, \cite{Allard_2023} and \cite{Allard_2024}. The dashed black curve corresponds to the critical opacity, $\kappa_{\rm crit}$. The dotted grey curve corresponds to $\kappa_{\rm R}$ from \cite{Guillot_1994}.}
    \label{fig:Jupiter_Na_opacity}
    \end{figure}


Using our opacity tables, we shall now determine the atmospheric compositions needed to develop a radiative zone on Jupiter. As was seen in the previous section, K and Na are the most important elements to determine the existence of a radiative zone. Hence, we show $\kappa_{\rm R}$ for Jupiter for varying elemental abundances of K and Na in Fig. \ref{fig:Jupiter_Na_opacity}. The abundances of all the other heavy elements are fixed at three times solar. The dashed black curve corresponds to the critical opacity, $\kappa_{\rm crit}$, of Jupiter, and is evaluated using Eq. (\ref{eq:critical_opacity}).
For comparison, we show $\kappa_{\rm R}$ from \cite{Guillot_1994}. As was expected, the presence of K, Na, and NaH has a significant impact on $\kappa_{\rm R}$ between $\sim 10^3-10^4$ bar. If their opacities are neglected, a radiative zone develops between $\sim 10^3-7\cdot 10^3$ bar (i.e. $\sim 1400-2100$ K). This region is comparable to the one identified by \cite{Guillot_1994}. However, the radiative zone in our study is shallower due to the inclusion of opacities from metal hydrides and oxides, which were not accounted for in \cite{Guillot_1994}. When the opacities of K, Na, and NaH are included, $\kappa_{\rm R}$ at the kilobar level increases by almost three orders of magnitude for solar abundances of K and Na. As the alkali abundance increases, the extent of the stable layer progressively shrinks at both ends until it disappears entirely and restores convection. Our opacities suggest that a significant alkali depletion of less than $\sim 10^{-3}$ times solar is required for a radiative zone to exist in Jupiter's molecular hydrogen envelope. 

Furthermore, we find that the radiative-convection boundary of Jupiter is located around 0.5 bar. However, this result is sensitive to the H$_2$-H$_2$ CIA opacity,  which below 200 K only covers wavelengths down to 4.167 $\mu$m (see Table \ref{table:BF_FF_opacity}). Hence, it is possible that the RCB may occur at even lower pressures.

In Fig. \ref{fig:Jupiter_mean_opacity_Metal_var}, we show the effect of all the other elements, except for K, Na, H, and He, on Jupiter's $\kappa_{\rm R}$. In this case, the atmospheres are highly depleted in K and Na ($10^{-5}$ times solar), while the abundances of other heavy elements are varied simultaneously. We find that, even with a high abundance of heavy elements, a radiative zone can still form deep in the atmosphere, provided there is sufficient alkali depletion.  This highlights again the crucial role of K, Na, and NaH in determining whether a radiative layer develops, as no other known gas species can prevent its formation. The main effect of non-alkali heavy elements on the deep radiative zone is to determine its extent. For a heavy element abundance of 0.1 times solar (consistent with interior models), the radiative layer extends from $\sim 5\cdot 10^2 - 10^4$ bar (i.e. $\sim 1200 - 2500$ K), as opposed to $\sim 10^3 - 7 \cdot 10^3$ bar (i.e. $\sim 1400 - 2100$ K) for a three times solar abundance suggested by Juno and Galileo observations.  

\subsubsection{Shallow radiative zone}

\begin{figure}[t]
   \centering
   \includegraphics[width=0.5\textwidth]{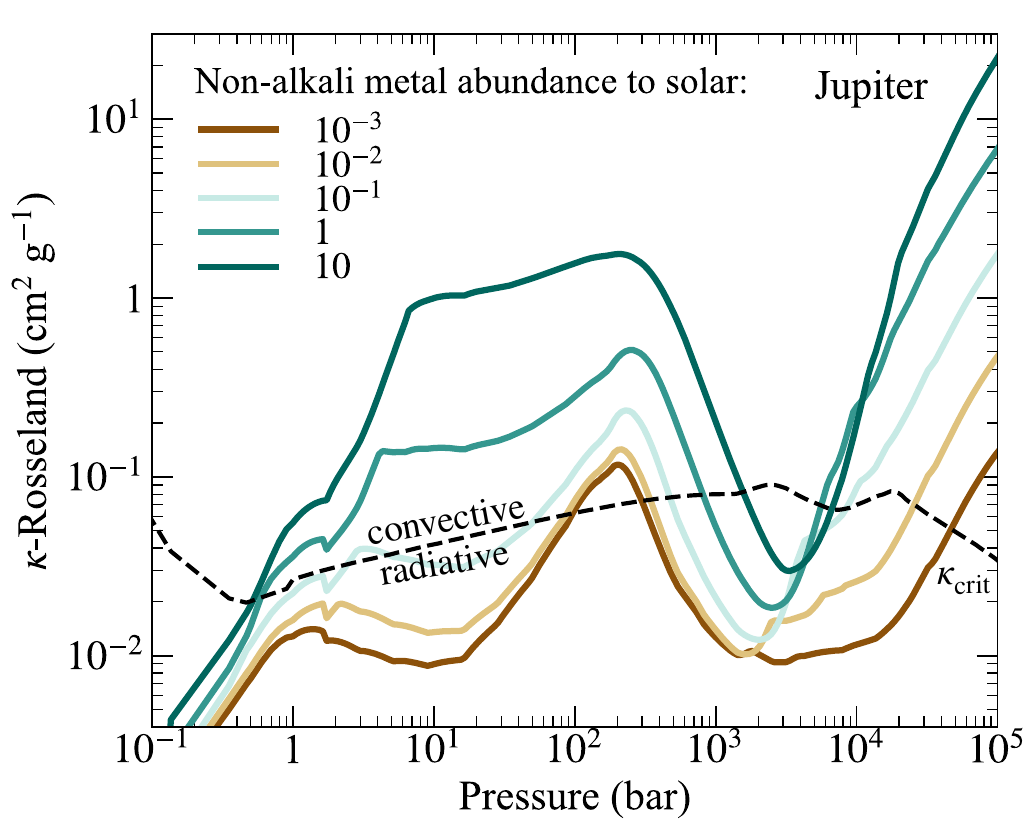}
   \caption{$\kappa_{\rm R}$ (solid coloured curves) as a function of pressure for atmospheres with varying heavy element abundances, but fixed Na and K abundances of $10^{-5}$ times solar. The opacities were calculated along a Jupiter profile. The dashed black curve corresponds to the critical opacity, $\kappa_{\rm crit}$.}
    \label{fig:Jupiter_mean_opacity_Metal_var}
    \end{figure}

In Fig. \ref{fig:Jupiter_mean_opacity_Metal_var}, we find that for low heavy element abundances, a second shallow radiative zone emerges between $\sim 3 - 100$ bar (i.e. $\sim 200-600$ K). This is related to a depletion in H$_2$O, NH$_3$, and CH$_4$, which are the main absorbers at low pressure and temperature. This radiative window was also identified by \cite{Guillot_1994} when removing the H$_2$O opacity. From Juno and Galileo measurements, we know that a depletion in NH$_3$ and CH$_4$ is unlikely at low pressures. The abundance in H$_2$O is more uncertain due to the subsolar abundance reported by the Galileo probe \citep{Wong_2004} and supersolar measurements from Juno \citep{Li_2024}. The work of \cite{Wong_2004} reports an O abundance of $0.289 \pm 0.096$ times solar abundance; however, we find that an abundance between $0.8$ and $1.1$ times solar is needed to align with the H$_2$O mixing ratio measurements from the Galileo probe. This discrepancy arises because the earlier study neglected the effect of rocks
and the sequestration of O atoms by these. At pressures around 10 kbar, MgO and SiO condense, and for a 0.289 solar O abundance, these species are abundant enough to almost completely remove O atoms from the atmosphere at lower pressures. As a result, the O abundance in \cite{Wong_2004} is incompatible with the Galileo H$_2$O measurement.

\begin{figure}[t]
   \centering
   \includegraphics[width=0.49\textwidth]{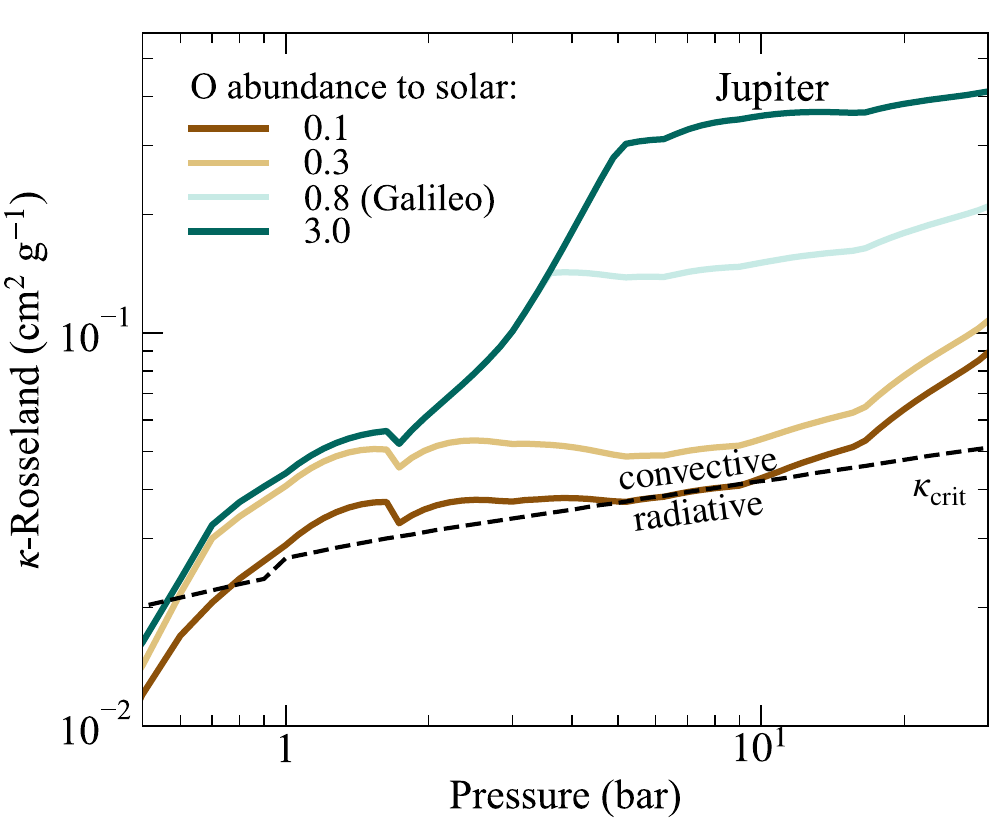}
   \caption{$\kappa_{\rm R}$ (solid coloured curves) at low pressures for Jupiter with varying O abundances. The abundance of C is fixed to 3.56 times solar, and that of N is fixed to 2.61 times solar. The dashed black curve corresponds to the critical opacity, $\kappa_{\rm crit}$}
    \label{fig:Jupiter_mean_opacity_Metal_var_2}
    \end{figure}

We re-calculated the opacities at low pressures using the measured  N and C abundances of 2.61 and 3.56 times solar, respectively \citep{Wong_2004, Li_2020, Guillot_2022}, while varying the O abundance. The abundance of K and Na has no effect in this region. The resulting $\kappa_{\rm R}$ are shown in Fig. \ref{fig:Jupiter_mean_opacity_Metal_var_2}. We find that with an O depletion of 0.1 times solar, the system is on the verge of preventing a radiative window, while for an abundance matching the Galileo measurement convection is restored. In fact, the temperature gradients of this region have been probed by the Galileo and Juno mission. During the in situ measurements, the Galileo probe found sub-adiabatic gradients between $0.5 - 20$ bar, which is thought to be a manifestation of mean molecular weight gradients due to the vertical variation in the water vapour abundance \citep{Magalh_2002}. These sub-adiabatic gradients are surprising, as our opacity models predict adiabatic gradients for an O-abundance matching the one measured by the Galileo probe. While we calculated mean molecular weight gradients for a case matching the O-abundance measured by the Galileo probe, we found them to be negligible and unable to produce sub-adiabatic gradients. One possible reason for this discrepancy is the use of equilibrium chemistry to describe the shallow layers, where temperatures are relatively low ($\lesssim 300$ K). Disequilibrium effects are expected in this region, which complicates the analysis. Furthermore, at pressures below $\sim$ 5 bar, water condensation is expected, releasing latent heat that can influence the local temperature.

\subsection{Conditions for a radiative zone on Saturn}

\begin{figure}[t]
   \centering
   \includegraphics[width=0.5\textwidth]{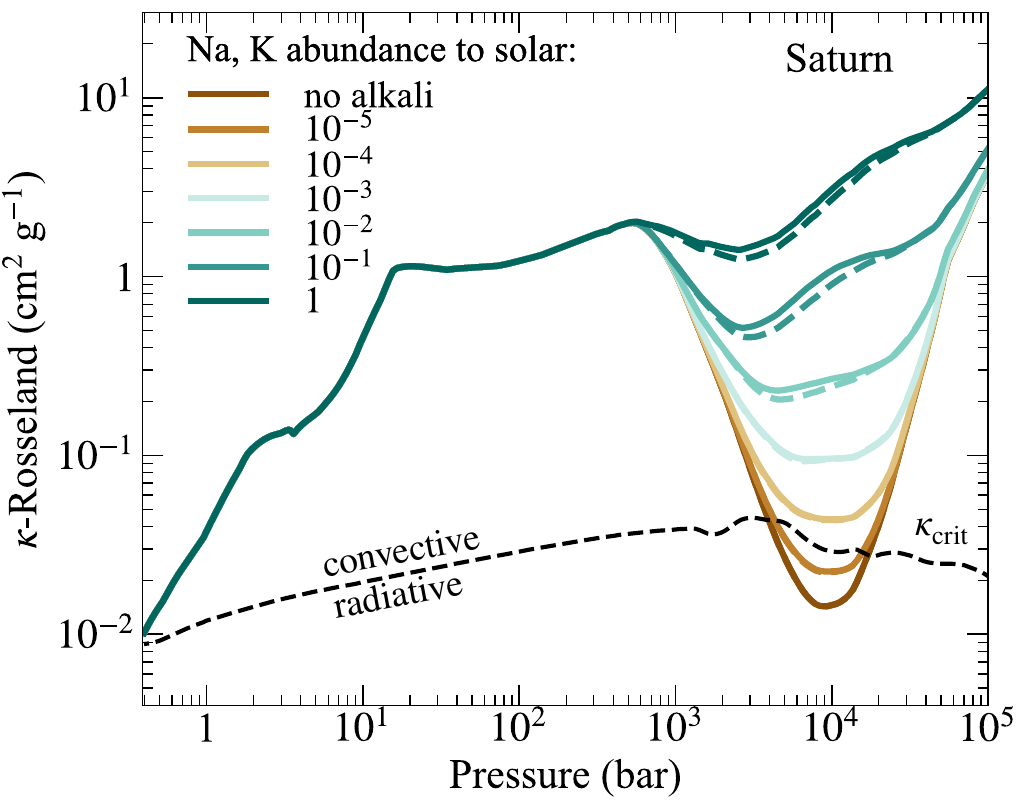}
   \caption{$\kappa_{\rm R}$ (solid and dashed coloured curves) as a function of pressure for atmospheres with varying and Na and K abundances calculated along a Saturn profile. The abundance of all other heavy elements is eight times solar. The solid curves were computed using the Na D and K I HWHM formulation from Eq. (\ref{eq:gamma_L_1}), while the dashed curves use the Na D and K I HWHM formulation from \cite{Allard_2016}, \cite{Allard_2019}, \cite{Allard_2023} and \cite{Allard_2024}. The dashed black curve corresponds to the critical opacity, $\kappa_{\rm crit}$. }
    \label{fig:Saturn_Na_opacity}
    \end{figure}

Having assessed the conditions for a radiative zone on Jupiter, we can now apply the same analysis to Saturn. Figure \ref{fig:Saturn_Na_opacity}
shows $\kappa_{\rm R}$ for Saturn with varying Na and K abundances, while the abundance of other heavy elements is fixed at eight times solar. Similar to Jupiter, a radiative zone forms only with sufficient alkali depletion. Non-alkali metals are unable to prevent the formation of the radiative layer and will only control its extent. Due to Saturn’s lower mass, the critical opacity, $\kappa_{\rm crit}$, is lower than Jupiter’s, implying that less Na and K are required to inhibit the development of a radiative zone. We find that an alkali abundance below $\sim 10^{-4}$ is required for a radiative layer to form, and it develops at higher pressures than on Jupiter, between $\sim 4 \cdot 10^3 - 2 \cdot 10^4$ bar.

The radiative-convection boundary on Saturn is located at $\lesssim$ 0.4 bar. The exact location cannot be determined because \texttt{FastChem} does not converge at temperatures below 100 K. As a result, we can only compute $\kappa_{\rm R}$ down to 0.4 bar for Saturn.

\subsection{Thermal profiles}

\begin{figure*}[t]
   \centering
   \includegraphics[width=1\textwidth]{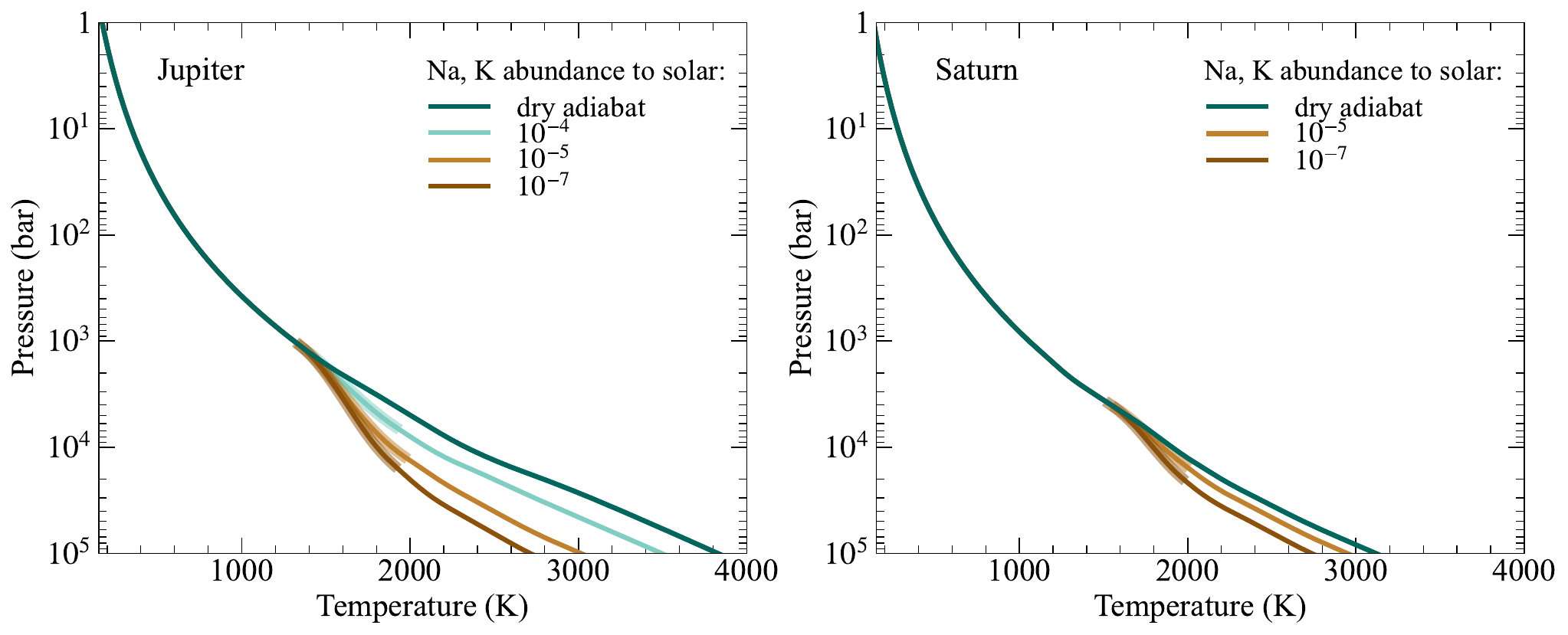}
   \caption{Temperature pressure profiles of Jupiter (left panel) and Saturn (right panel) for different K and Na abundances. The parts where radiative zones exist are indicated as the shaded regions. The adiabatic models are given as green curves.}
    \label{fig:tp_profiles}
    \end{figure*}

One of the main effects of a radiative zone in the deep atmosphere of a planet is to introduce deviations from an adiabatic TP profile. In Fig. \ref{fig:tp_profiles}, we show TP profiles of Jupiter and Saturn in the presence of a radiative zone (shown as shaded regions). This was achieved by sufficiently depleting the atmosphere in K and Na, while keeping the abundance of the other heavy elements fixed at three and eight times solar for Jupiter and Saturn, respectively. As was anticipated, a radiative zone results in a cooler interior of the planet, given its subadiabatic nature ($\nabla_{\rm rad} < \nabla_{\rm ad}$). 
For Jupiter, a significant depletion in alkali results in a temperature at the bottom of the radiative zone that is at most $\sim 25\%$ smaller than the one of the dry adiabatic model. The temperature difference is slightly less than the $30\%$ found in \cite{Guillot_1994b}. The origin of this is most likely the inclusion of more opacity sources in this work, which increases the  temperature gradients in the radiative zone. We expect that the temperature difference propagates to the centre of the planet, as the convective gradients of the dry adiabat and the radiative zone model are similar at deeper levels. As the alkali abundance increases, the temperature difference decreases as a result of the larger temperature gradient in the radiative zone.  

For Saturn, a significant depletion in alkali leads to a smaller temperature difference at the bottom of the radiative zone relative to a dry adiabat model. This is related to the fact that the radiative zone is located at deeper levels in Saturn, which reduces its effect. In addition, the heavy element abundance is larger on Saturn, which increases the temperature gradient in the radiative zone. We find that the radiative zone can lead to an interior which is at most $\sim 10 \%$ colder than the adiabatic model, which again is slightly less than the 15$\%$ temperature decrease reported by \cite{Guillot_1994b}.

Lowering the planet's interior temperature has significant implications, including shifting the locations of key phase transitions, such as H/He separation \citep{Salpeter_1973, Stevenson_1979} and the transition to metallic hydrogen \citep{Sano_2011, Loubeyre_2012}. For a cooler interior, these transitions occur deeper within the planet. Another consequence of a cooler interior is to increase the core density. Since the mean density of the planet is constrained by observations, an increase in the central density must be balanced by a decrease in density in the gaseous envelope. In other words, the atmospheric heavy element abundance, $Z_{\rm atm}$, is lower for interior models that include a radiative zone compared to adiabatic models.  

The presence of a radiative zone also modifies the internal entropy of the planet. By definition, for a fully convective interior, without any phase transitions, the specific entropy, $S$, is constant throughout the planet. However, in a radiative zone, $S$ will change due to its non-adiabatic behaviour. We find that $S$ decreases from the top to the bottom of the radiative zone by a magnitude of up to $\Delta S \approx - 4 \cdot 10^7 \ \rm erg \ K^{-1} g^{-1}$ for Jupiter and $\Delta S \approx - 2 \cdot 10^7 \ \rm erg \ K^{-1} g^{-1}$ for Saturn. The magnitude, $\Delta S$, decreases as the abundance of K and Na increases until the medium becomes fully convective again.  

\section{Discussion} \label{sec:Discussion}

\subsection{Half width at half maximum and cut-off of K I and Na D lines} \label{sec:cutoff_HFWHM}

\begin{figure}
   \centering
   \includegraphics[width=0.47\textwidth]{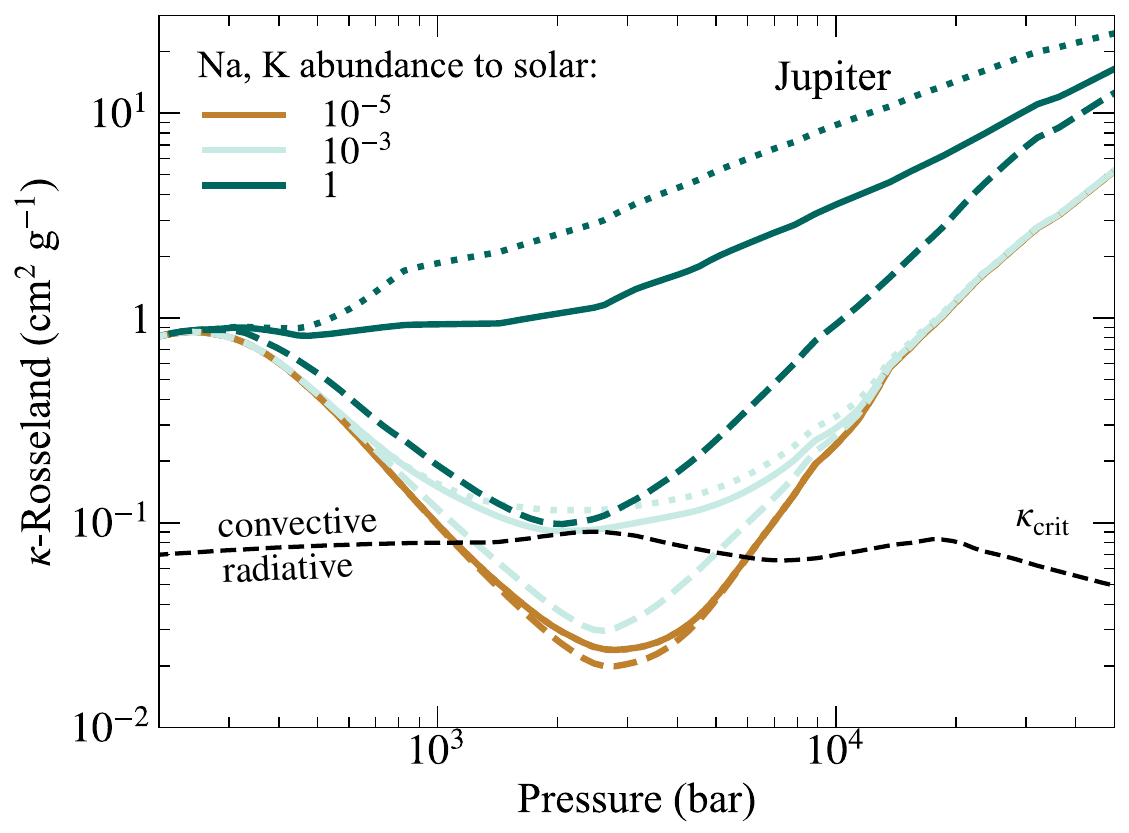}
   \caption{$\kappa_{\rm R}$ of Jupiter for varying Na and K abundances and adopting different line wing cut-off values for the K I and Na D features. The solid lines represent the baseline cut-off value of 4500 cm$^{-1}$, the dotted lines use a cut-off value of 10000 cm$^{-1}$, and the dashed lines use a cut-off value of 1000 cm$^{-1}$. The dashed black curve corresponds to the critical opacity, $\kappa_{\rm crit}$.}
    \label{fig:Jupiter_cutoffs}
    \end{figure}

In this paper, we have used the calculations from  \cite{Allard_2016} and \cite{Allard_2019} to model the K I and Na D resonance lines (see Sect. \ref{sec:atomic_lines}). At high pressure, we switched to Voigt profiles due to the lack of applicable theories and experiments. In Jupiter and Saturn, K and Na only become abundant at pressures $\gtrsim 1000$ bar, at which point we modelled their resonance lines as Voigt profiles with
a Lorentzian HWHM based on Van der Waals broadening parameters and a cut-off value of 4500 cm$^{-1}$ for the line wings. Alternative formulations for the Lorentzian HWHM of the resonance lines exist, which depend on the assumed potential energy curve of the two colliding partners. To assess the impact on our results, we explored HWHM formulations from various sources: \cite{Allard_2016} for the K-H$_2$ potential, \cite{Allard_2024} for the K-He potential, \cite{Allard_2019} for the Na-H$_2$ potential, and \cite{Allard_2023} for the Na-He potential. We find that the K and Na resonance feature is narrower when using the Allard HWHM treatment.  However, the differences in $\kappa_{\rm R}$ are negligible, and our conclusions remain unchanged regardless of the Lorentzian HWHM formulation used (see Fig. \ref{fig:Jupiter_Na_opacity} and \ref{fig:Saturn_Na_opacity} dashed coloured curves). 

The choice of line wing cut-off value is significantly more important. \cite{Baudino_2017} and \cite{Chubb_2020} recommend a cut-off value of 4500 cm$^{-1}$, although this is not based on any specific theoretical justification, but rather on the observation that the K I and Na D line wings extend significantly from their core. $\kappa_{\rm R}$ of Jupiter assuming different line wing cut-offs are shown in Fig. \ref{fig:Jupiter_cutoffs}. We find that increasing the cut-off value from 4500 cm$^{-1}$ to 10000 cm$^{-1}$ (dotted curve) has no notable effect on $\kappa_{\rm R}$, and, as before, an alkali depletion of less than $10^{-3}$ times solar is required to develop a radiative zone. In contrast, reducing the cut-off value to 1000 cm$^{-1}$ (dashed curve) substantially decreases $\kappa_{\rm R}$. In this scenario, a radiative layer can develop even with alkali abundances close to solar levels. While it is unlikely that the Na D and K I line wings only extend to 1000 cm$^{-1}$, given their behaviour at lower pressures, this example highlights the importance of improving our understanding of their profiles at high pressures. We emphasise that regardless of the cut-off value, a Na abundance above solar should still be sufficient to prevent a radiative zone due to the opacity of NaH.

\subsection{Constraints on alkali abundance}
While we have determined how much K and Na is needed to develop a deep radiative layer on Jupiter and Saturn, the true abundance of these elements is unclear. Since they are expected to condense out of the atmosphere around 800 K, they should not be directly observable in cooler planets like Jupiter and Saturn. The resonance feature Na D has been observed in ‘transmission’ spectra of Jupiter \citep{Montanes_Rodriguez_2015}, although this feature is likely related to the continuous volcanic outgassing from the galilean moon Io \citep{Lellouch_2003, MENDILLO_2004} or cometary impacts \citep{Noll_1995}. Hence, these observations should not be representative of the alkali abundance at deeper levels.

Recent brightness temperature and limb darkening observations by the MWR instrument on Juno attempted to constrain the abundance of K and Na, as a proxy for the free electron abundance at the kilobar level. The initial results indicate a high depletion in alkali in the range of $10^{-5}$ to $10^{-2}$ times solar \citep{Bhattacharya_2023}. This level of depletion would potentially allow for the existence of a stable radiative layer in Jupiter, based on the calculations performed in this study. However, this study neglected the importance of anions, which act as a sink of free electrons, and a reassessment of the analysis suggested a higher abundance for both K and Na of order $10^{-1}$ times solar \citep{Aglyamov_2024}. Although there still seems to be a subsolar abundance in K and Na on Jupiter, our results and these measurements suggest that developing a radiative zone on the planet at this time would be challenging. The origin of such a depletion is unclear and could be explained through an unconventional planet formation mechanism, where Jupiter accreted a subsolar abundance of alkali metals. Alternatively, it could imply that the planet's interior is not well mixed, such that most alkalis are locked in the core. A diluted core, as has been proposed by recent interior models \citep{Wahl_2017, Nettelmann_2021, Miguel_2022}, can reduce the efficiency of convective mixing, and it has been shown that for Jovian conditions it has an erosion lifetime exceeding Jupiter's age \citep{Fuentes_2024}. We emphasise that the alkali measurements provided by the MWR observations are interpreted as global abundances on Jupiter and are expected to vary spatially due to the complexity of the planet's atmosphere. Alternative measurements of Jupiter's atmospheric alkali levels would be invaluable in confidently ruling out the existence of a radiative layer.
 
Alkali metals have been detected in hot gas giant exoplanets. An indirect measurement includes Kepler-7b,  where an alkali depletion of $10^{-1} - 10^{-2}$ was inferred based on the planets high albedo \citep{Demory_2011}. Another subsolar abundance in alkali was observed in the transit spectra of HAT-P-1b, with $\rm [Na/H] = -0.77^{+0.50}_{-0.44}$ and $\rm [K/H] < -1.25$ \citep{Chen_2022}. While these measurements suggest alkali depletion on gas giants, solar to super-solar measurements also exist. Examples include WASP-76b with [Na/H]$ = 0.649^{+0.511}_{-0.902}$ \citep{Fu_2021} or WASP-96b with [Na/H]$ = 1.32^{+0.36}_{-0.48}$ \citep{Nikolov_2018, Nikolov_2022}. Given the discrepancy in alkali abundances and the small sample size of their detection in gas giant exoplanets, we should refrain from drawing any conclusions about the alkali abundance on Jupiter or Saturn based on exoplanet observations. However, we note that repeating these calculations to giant exoplanets with known alkali abundances would be of great interest.

\subsection{Impact of cloud opacities}

\begin{figure}
   \centering
   \includegraphics[width=0.47\textwidth]{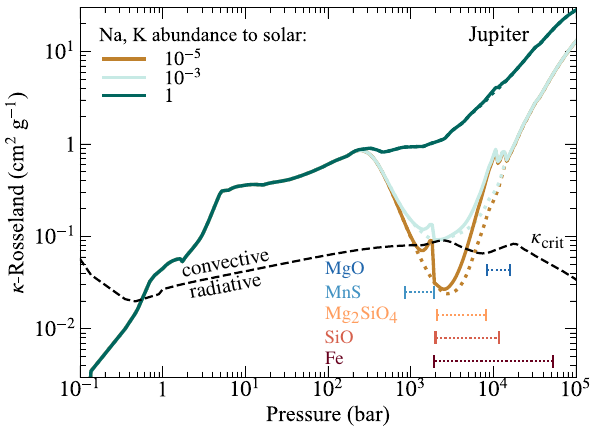}
   \caption{Solid curves give $\kappa_{\rm R}$ of Jupiter for varying Na and K abundances and including the opacities from the condensates Fe, MgO, Mg$_2$SiO$_4$, MnS, and SiO. The abundance of all other heavy elements was set to three times solar. The regions where each condensate contributes to $\kappa_{\rm R}$ are illustrated by the bars. The dotted curves give $\kappa_{\rm R}$ without the contribution from condensate opacities. The dashed black curve corresponds to the critical opacity, $\kappa_{\rm crit}$.}
    \label{fig:Jupiter_clouds}
    \end{figure}

As is pointed out in Sect. \ref{sec:Atmospheric_Composition}, heavy elements condense at high temperatures and can form cloud decks. Molecules that condense between $10^3 - 10^4$ bar for Jupiter and Saturn conditions are, for example, Fe, MgO, Mg$_2$SiO$_4$, MnS, and SiO. This represents an additional source of opacity that also extends to wavelengths $< 1\mu \rm m$. We shall estimate here the impact of these five condensates on $\kappa_{\rm R}$.  

Modelling cloud formation and their opacity involves a description of condensation, coagulation and coalescence, sedimentation, and eddy mixing. Since at high temperatures ($> 1000$ K), eddy mixing timescales are expected to be much slower than the ones of other processes, the structure of clouds will be similar to that determined by the condensates vapour pressure (i.e. the output of \texttt{FastChem}). We simplified the coagulation and sedimentation process using the \cite{Ackerman_2001} model, which allows one to parametrise the mean radius and number density of cloud particles in a layer using two parameters, $f_{\rm rain}$ (ratio of mass-weighted sedimentation velocity to convective velocity scale), and $\sigma_{\rm g}$ (geometric standard deviation in lognormal size distributions of condensates). We defined $f_{\rm rain} = 3$ and $\sigma_{\rm g} = 2$, which results in mean radii ranging from $\sim 1 - 3 \ \mu \rm m$. We computed the absorption and scattering cross-sections of condensates using the code \texttt{LX-MIE} \citep{Kitzmann_2017} and the compilation of the optical constants therein. Figure \ref{fig:Jupiter_clouds} shows $\kappa_{\rm R}$ of Jupiter, including the contributions from condensates and assuming a heavy element abundance of three times solar. The dotted coloured bars represent the extent of the cloud deck for each condensate. At low K and Na abundances, sharp peaks in $\kappa_{\rm R}$ appear, corresponding to the base of cloud decks where condensates have settled, creating regions of higher opacity. The first peak, around $2 \cdot 10^3$ bar, is due to MnS condensates, while the second and third peaks are caused by SiO and MgO condensates, respectively. As the abundance of K and Na increases, these cloud-related features disappear, as alkali opacities dominate. We find that the MnS cloud is able to elevate $\kappa_{\rm R}$ above the critical value, $\kappa_{\rm crit}$, shrinking the extent of the radiative layer. The cloud base of the other condensates resides at higher pressure, $\gtrsim 10^4$ bar, and as a result they are unable to prevent the radiative zone. Varying the heavy element abundance alters the location of these cloud bases; generally, a higher abundance raises the vapour saturation pressure, pushing the cloud deck to higher pressures. Conversely, decreasing the abundance shifts the cloud base into the radiative zone but also reduces its opacity. Hence, the condensates considered here are incapable of fully preventing the radiative zone. In this analysis, we restricted ourselves to condensates for which optical properties were available. We found that condensates of Cu, Cr$_2$O$_3$, CaTiO$_3$, CaSiO$_3$, and CaMgSi$_2$O$_6$ form cloud bases in the region of the radiative zone, although their mixing ratios are smaller than that of MnS at three times solar metallicity. However, calculating $\kappa_{\rm R}$ with the contribution of their opacities would be of great interest.

\subsection{Validity of Schwarzschild criterion}
As is outlined in Sect. \ref{sec:heat_transport}, we employed the Schwarzschild stability criterion to determine the dominant heat transport mechanism. However, this criterion applies to homogeneous and dry conditions, and the presence of condensation can violate these assumptions. When a species condenses, this produces a vertical gradient in the mixing ratio of that species in the gas phase,  with the mixing ratio increasing with depth. As a result, when condensation takes place, the mean molecular weight of the atmosphere, denoted as 
$\mu$, can increase at greater depths  (i.e. $\nabla_\mu = \frac{d\rm ln \ \mu}{d\textrm{ln} \ p} > 0$), thereby stabilising the medium against convection. This is known as the Ledoux stability criterion \citep{Ledoux_1947},

\begin{equation}
    \nabla_{\rm rad} \leq \nabla_{\rm ad} + \nabla_\mu.
\end{equation}

\noindent Hence, it is possible for a radiative layer to exist for a superadiabatic gradient. The next consequence of condensation is that, within a convective layer, the temperature profile follows the ‘moist’ adiabat, $\nabla_{\rm ad}^{\star}$, instead of the dry adiabat, $\nabla_{\rm ad}$. This change occurs due to the release of latent heat during condensation, which contributes to a positive buoyancy and facilitates convection; that is, $\nabla_{\rm ad}^{\star} < \nabla_{\rm ad}$.

Since condensation of refractory materials occurs in the region where a radiative layer may exist in Jupiter and Saturn, it is reasonable to question the use of the Schwarzschild criterion in this study. We calculated mean molecular weight gradients for atmospheres assuming different abundances of metals, and found that they never exceeded a value of $0.2 \nabla_{\rm ad}$ at high pressures. While this introduces a change in the critical opacity, $\kappa_{\rm crit}$, its effect is negligible and will not change the results of this work. At lower pressures, in the region where H$_2$O condenses, the effect of mean molecular weight gradients is more pronounced and is known to stabilise Jupiter's atmosphere in the $5 \ \mu$m hotspot \citep{Magalh_2002}.  

We have also estimated the moist adiabat caused by the condensation of MgO at high pressures. We chose MgO due to its large known latent heat of vapourisation $L = 8.27 \ \rm kJ \ g^{-1}$ \citep{Mahmoud_2017} compared to other condensing species with known values (e.g. $L = 6.34 \ \rm kJ \ g^{-1}$ for Fe, see \citealt{Zhang_2011}). Following Eq. (11) in \cite{Leconte_2017}, we find that the fractional difference between the dry and moist adiabat is of order $10^{-8}$, which justifies the use of the dry adiabat in this work. 
   

\section{Conclusions} \label{sec:Conclusions}

We have calculated $\kappa_{\rm R}$ for the molecular hydrogen envelopes of Jupiter and Saturn. Our opacity tables are the most comprehensive to date, covering pressures up to $10^5$ bar and incorporating the most abundant molecules found in gas giants, along with contributions from free electrons, various metal hydrides and oxides, and atomic species. We investigated different atmospheric composition and determined the conditions needed to develop a radiative zone at the kilobar level in Jupiter and Saturn today. 

We found that K, Na, and NaH are the only species capable of restoring convection in the kilobar regime. The main effect of other opacity species is to control the width of a radiative zone. For significant depletions in the elemental abundances of K and Na, a radiative zone develops on Jupiter between $\sim 10^3- 7 \cdot 10^3$ bar, while on Saturn it exists deeper in the envelope between $\sim 4 \cdot 10^3- 2 \cdot 10^4$. For K and Na abundances exceeding $\sim 10^{-3}$ times solar on Jupiter and $\sim 10^{-4}$ times solar on Saturn, convection is restored. The radiative zone identified in this study is too shallow to control the depth of Jupiter's zonal winds as is suggested by \cite{Christensen_2024} and \cite{Moore_2022}. At pressures $\gtrsim 10^4$ bar, the radiative opacities are more than two orders of magnitude too large too allow for stable radiative layers at greater depths. However, this does not rule out the possibility of stable layers at deeper levels, as conductive opacities were not calculated in this study. In a highly conductive fluid, heat transport can occur through conduction, potentially resulting in stable conductive layers.

We determined new thermal profiles for Jupiter and Saturn, assuming sufficient depletions of K and Na to form a radiative layer. Our results show that the presence of a radiative zone lowers Jupiter's interior temperature by up to $25\%$ compared to an adiabatic model, while Saturn's interior temperature decreases by up to $10\%$. The cooler interior has significant implications for the inferred metallicity of the planet's gaseous envelope in interior models. Additionally, a radiative zone decreases the specific entropy by up to $\Delta S \approx -4 \cdot 10^7 \rm erg \ K^{-1} \ g^{-1}$ for Jupiter and $\Delta S \approx -2 \cdot 10^7 \rm erg \ K^{-1} \ g^{-1}$ for Saturn. 

We emphasise that these results are strongly influenced by the cut-off value for the line wings of the K I and Na D resonance lines, as is discussed in Sect. \ref{sec:cutoff_HFWHM}. Developing a theory for these lines in the kilobar regime would enhance the accuracy of this study. Furthermore, it is important to mention that molecules without available line lists could influence our results. For instance, potassium hydride (KH), which forms at conditions similar to NaH, likely has absorption features below $1 \ \mu \rm m$, and its absence in our opacity tables could affect the results.

In summary, we show that a substantial depletion in K and Na is needed to develop a radiative zone in the deep atmosphere of Jupiter and Saturn. Current estimates of their abundances on Jupiter suggest subsolar levels, which would restore convection at the present day.

We emphasise that, to either confidently rule out the possibility of a radiative zone or determine if it can resolve the conflict between interior models and observations, several critical steps are required. First, new methods of probing alkali should be developed to gather additional measurements and confirm the current constraints on their abundance. Second, we need to advance the theory describing the K I and Na D line profiles when perturbed by H$_2$ and He, particularly at higher pressures. Finally, understanding the dynamics and evolution of the radiative zone is essential. This includes exploring molecular diffusion across the layer and investigating its properties and existence during earlier stages of planetary evolution.

\vspace{0.5cm}
\begin{spacing}{0.81}
\noindent{\tiny \emph{Data Availability.} The radiative opacities used in this study, along with $\kappa_{\rm R}$ tables for Jupiter and Saturn, are available at \url{https://doi.org/10.5281/zenodo.14507606}.}
\end{spacing}

\begin{acknowledgements}
      We thank the referee for valuable comments which helped improve the manuscript.
      This project has received funding from the European Research Council (ERC) under the European Union’s Horizon 2020 research and innovation programme (grant agreement no. 101088557, N-GINE). This publication is part of the project ENW.GO.001.001 of the research programme “Use of space infrastructure for Earth observation and planetary research (GO), 2022-1” which is (partly) financed by the Dutch Research Council (NWO). TG benefited from support from CNES as part of the Juno mission.
\end{acknowledgements}

%
%

\bibliographystyle{aa} 
\bibliography{references.bib} 


\begin{appendix}

\onecolumn
\section{Tables}

\begin{table}[h!]
\small
\centering
 \caption{Molecular opacities used in this work. $^*$ super-line lists were used for H$_2$O and NH$_3$.}
 \begin{tabular}{c @{\hskip 0.3cm } c @{\hskip 0.3cm } c @{\hskip 0.3cm } c @{\hskip 0.3cm } c @{\hskip 0.3cm } c} 
 \hline
 \hline
Molecule  & Temperature (K) & Pressure (bar) & Wavelength ($\mu \rm m$) & Line List Name &  References \\ [0.01ex] 
  \noalign{\smallskip}
    \hline
    \noalign{\smallskip}
    AlH & $100-5000$ & $10^{-1} - 10^5$  & $0.407 - 200$ & AloHa &\textrm{\cite{Sergei_2023}}  \\
    CaH & $100-5000$ & $10^{-1} - 10^5$  & $0.335 - 200$ & XAB &\textrm{\cite{Owens_2022a}}  \\
    CaO & $100-5000$ & $10^{-1} - 10^5$  & $0.400 - 200$ & VBATHY &\textrm{\cite{Yurchenko_2016}}  \\
    CH & $100-5000$ & $10^{-1} - 10^5$  & $0.255 - 200$ & MoLLIST &\textrm{\cite{Masseron_2014, Bernath_2020}}  \\
    CH$_4$ & $50-2500$ & $10^{-8} - 10^3$  & $0.746 - 200$ & HITEMP2020 &\textrm{\cite{Rothman_2010, Hargreaves_2020}}  \\
    CO & $100-5000$ & $10^{-1} - 10^5$  & $0.455 - 200$ & Li2015 &\textrm{\cite{Li_2015, Somogyi_2021}}  \\
    CO$_2$ & $50-4500$ & $10^{-8} - 10^3$  & $0.500 - 200$ & UCL-4000 &\textrm{\cite{Yurchenko_2020}}  \\
    CP & $100-3000$ & $10^{-1} - 10^5$  & $0.661 - 4.386$ & MoLLIST &[1] \\
    CrH & $100-3000$ & $10^{-1} - 10^5$  & $0.667 - 1.376$ & MoLLIST & [2]  \\
    FeH & $100-5000$ & $10^{-1} - 10^5$  & $0.667 - 23$ & MoLLIST &\textrm{\cite{Dulick_2003, Bernath_2020}} \\
    H$_2$ & $100-5000$ & $10^{-1} - 10^5$  & $0.278 - 200$ & RACPPK &\textrm{\cite{Roueff_2019}}\\
    H$_2$O & $100-5000$ & $10^{-1} - 10^5$  & $0.243 - 200$ & POKAZATEL$^*$ &\textrm{\cite{Polyansky_2018}}\\
     H$_2$S & $100-3000$ & $10^{-1} - 10^5$  & $0.286 - 200$ & AYT2 &\textrm{\cite{Azzam_2016, Chubb_2018}}\\
    HCl & $100-5000$ & $10^{-1} - 10^5$  & $0.494 - 200$ & HITRAN-HCl &\textrm{\cite{Gordon_2017}}\\
    HF & $100-5000$ & $10^{-1} - 10^5$  & $0.31 - 200$ & Coxon-Hajig & [3]\\
    MgH & $100-5000$ & $10^{-1} - 10^5$  & $0.338 - 200$ & XAB &\textrm{\cite{Owens_2022a}}\\
    MgO & $100-5000$ & $10^{-1} - 10^5$  & $0.270 - 200$ & LiTY &\textrm{\cite{Li_2019}}\\
    N$_2$ & $100-5000$ & $10^{-1} - 10^5$  & $0.179 - 200$ & WCCRMT & [4]\\
    NaCl & $100-3000$ & $10^{-1} - 10^5$  & $4.069 - 200$ & Barton &\textrm{\cite{Barton_2014}}\\
    NaH & $100-5000$ & $10^{-1} - 10^5$  & $0.311 - 200$ & Rivlin &\textrm{\cite{Rivlin_2015, Chubb_2020}}\\
    NH$_3$ & $100-2000$ & $10^{-1} - 10^5$  & $0.500 - 200$ & CoYuTe$^*$ &\textrm{\cite{Derzi_2015, Coles_2019}}\\
    PH$_3$ & $50-2900$ & $10^{-8} - 10^3$  & $0.100 - 200$ & SAITY &\textrm{\cite{Silva_2014}}\\
    PS & $100-5000$ & $10^{-1} - 10^5$  & $0.270 - 200$ & POPS &\textrm{\cite{Prajapat_2017}}\\
    SiH & $100-5000$ & $10^{-1} - 10^5$  & $0.313 - 200$ & SiGHTLY &\textrm{\cite{Yurchenko_2017}}\\
    SiH$_4$ & $50-1900$ & $10^{-8} - 10^3$  & $2 - 200$ & OY2T &\textrm{\cite{Owens_2017}}\\
    SiO & $100-5000$ & $10^{-1} - 10^5$  & $0.139 - 200$ & SiOUVenIR &\textrm{\cite{Yurchenko_2021}}\\
    SO & $100-5000$ & $10^{-1} - 10^5$  & $0.222 - 200$ & SOLIS &\textrm{\cite{Brady_2023}}\\
    TiH & $100-4800$ & $10^{-1} - 10^5$  & $0.417 - 2.156$ & MoLLIST &\textrm{\cite{Burrows_2005, Bernath_2020}}\\
    TiO & $100-5000$ & $10^{-1} - 10^5$  & $0.333 - 200$ & Toto &\textrm{\cite{McKemmish_2019}}\\
    VO & $100-5000$ & $10^{-1} - 10^5$  & $0.286 - 200$ & VOMYT &\textrm{\cite{McKemmish_2016}}\\
        
  \hline
\end{tabular}
 \label{table:molecule_opacity}
 \begin{flushleft}
\tiny{[1]: \textrm{\cite{Ram_2014, Bernath_2020, Qin_2021}} [2]: \textrm{\cite{Burrows_2002, Chubb_2018, Bernath_2020}} [3]: \textrm{\cite{Li_2015, Coxon_2015, Somogyi_2021}} [4]: \textrm{\cite{Shemansky_1969, Western_2017, Western_2018, Jans_2024}}} 
\end{flushleft}
 
\end{table}

\begin{table}[h]
\small
\centering
 \caption{Collision-induced absorption used in this work.}
 \begin{tabular}{c @{\hskip 1.25cm }  c @{\hskip 1.25cm } c @{\hskip 1.25cm } c @{\hskip 1.25cm } c @{\hskip 0.1cm } c} 
 \hline
 \hline
Species Name  & Temperature (K) & Pressure (bar) & Wavelength ($\mu \rm m$) & Reference \\ [0.01ex] 
  \noalign{\smallskip}
    \hline
    \noalign{\smallskip}
    H$_2$-H$_2$ & $200-3000$ & $10^{-1} - 10^5$  & $1 - 200$ & \textrm{\cite{Abel_2012}}  \\
    & $100-200$ & $10^{-1} - 10^5$  & $4.167 - 200$ & \textrm{\cite{Fletcher_2018}}  \\
    H$_2$-H & $1000-2500$ & $10^{-1} - 10^5$ & $1 - 100$ &\textrm{\cite{Gustafsson_2003}}\\
    H$_2$-He & $200-5000$ & $10^{-1} - 10^5$  & $0.500 - 200$ & \textrm{\cite{Abel_2011}} \\
    H$_2$-CH$_4$ & $100-400$ & $10^{-1} - 10^5$ &  $5.139 - 200$ & \textrm{\cite{Borysow_1986}}  \\

  \hline
\end{tabular}
 \label{table:CIA_opacity}
\end{table}

\begin{table}[h]
\small
\centering
 \caption{Bound-free and free-free absorptions considered in this work.}
 \begin{tabular}{c @{\hskip 5.cm }  c @{\hskip 5.cm }  c @{\hskip 0.1cm } c} 
 \hline
 \hline
Reaction & Wavelength ($\mu$m) & Reference \\ [0.01ex] 
  \noalign{\smallskip}
    \hline
    \noalign{\smallskip}
    $\textrm{H}_2 + \textrm{e}^- + \textrm{h}\nu \xrightarrow{} \textrm{H}_2 +  \textrm{e}^-$ & $0.351 - 15.188$ & \textrm{\cite{Bell_1980}}  \\
     $\textrm{H} + \textrm{e}^- + \textrm{h}\nu \xrightarrow{} \textrm{H} +  \textrm{e}^-$ & $0.182 - 200$ &\textrm{\cite{John_1988}}  \\
     $\textrm{H}^-  + \textrm{h}\nu \xrightarrow{} \textrm{H} +  \textrm{e}^-$ & $0.125 - 1.642$ & \textrm{\cite{John_1988}}  \\
     $\textrm{He} + \textrm{e}^- + \textrm{h}\nu \xrightarrow{} \textrm{He} +  \textrm{e}^-$ & $0.506 - 200$ & \textrm{\cite{John_1994}}  \\
  \hline
\end{tabular}
 \label{table:BF_FF_opacity}

\end{table}

\end{appendix}

\end{document}